\documentclass[twocolumn]{aastex631}

\usepackage{natbib}
\usepackage{booktabs}
\bibpunct{(}{)}{;}{a}{}{,}

\usepackage{amsmath,amssymb,amsfonts}
\usepackage{graphicx}
\usepackage{txfonts}
\usepackage{xcolor}

\usepackage{CJK}

\DeclareRobustCommand{\ion}[2]{\textup{#1\,\textsc{\lowercase{#2}}}}

\begin{document}
\begin{CJK*}{UTF8}{gbsn}

\title{High-resolution transmission spectroscopy of ultra-hot Jupiter WASP-33b with NEID}

\correspondingauthor{Guo Chen}
\email{guochen@pmo.ac.cn}

\author[0009-0001-5682-5015]{Yuanheng Yang (杨远恒)}
\affiliation{CAS Key Laboratory of Planetary Sciences, Purple Mountain Observatory, Chinese Academy of Sciences, Nanjing 210023, China}
\affiliation{School of Astronomy and Space Science, University of Science and Technology of China, Hefei 230026, China}

\author[0000-0003-0740-5433]{Guo Chen (陈果)}
\affiliation{CAS Key Laboratory of Planetary Sciences, Purple Mountain Observatory, Chinese Academy of Sciences, Nanjing 210023, China}
\affiliation{CAS Center for Excellence in Comparative Planetology, Hefei 230026, China}

\author[0000-0002-7846-6981]{Songhu Wang}
\affiliation{Department of Astronomy, Indiana University, Bloomington, IN 47405, USA}

\author[0000-0001-9585-9034]{Fei Yan (严飞)}
\affiliation{Department of Astronomy, University of Science and Technology of China, Hefei 230026, China}



\begin{abstract}

We report an attempt to detect molecular and atomic species in the atmosphere of the ultra-hot Jupiter WASP-33b using the high-resolution echelle spectrograph NEID with a wavelength coverage of 380$-$930 nm. By analyzing the transmission spectrum of WASP-33b using the line-by-line technique and the cross-correlation technique, we confirm previous detection of H$\alpha$, H$\beta$, H$\gamma$, and \ion{Ca}{ii} infrared triplets. We find no evidence for a significant day-to-night wind in WASP-33b, taking into account the effects of stellar pulsations using a relatively novel GP method and poorly constrained systemic velocity measurements. We also detect the previously reported pre-transit absorption signal, which may be a pulsation mode induced by the planet. Combined with previous CARMENES and HARPS-N observations, we report the non-detection of TiO, \ion{Ti}{i}, and \ion{V}{i} in the transmission spectrum, while they were already detected in the dayside atmosphere of WASP-33b. This implies a difference in the chemical compositions and abundances between the dayside and terminator atmospheres of WASP-33b, and certainly requires further improvements in the sensitivity of the detection methods.

\end{abstract}


\keywords{\href{http://astrothesaurus.org/uat/487}{Exoplanet atmospheres (487)}; 
\href{http://astrothesaurus.org/uat/2021}{Exoplanet atmospheric composition (2021)}; 
\href{http://astrothesaurus.org/uat/2133}{Transmission spectroscopy (2133)};
\href{http://astrothesaurus.org/uat/2096}{High resolution spectroscopy (2096)}; 
\href{http://astrothesaurus.org/uat/753}{Hot Jupiters (753)}; 
\href{http://astrothesaurus.org/uat/498}{Exoplanets (498)};
\href{http://astrothesaurus.org/uat/509}{Extrasolar gaseous giant planets (509)}
}


\section{Introduction}
\end{CJK*}

	Ultra-hot Jupiters (UHJs) are gas giant planets with high dayside temperatures \citep[$T_\text{day}$ $\ge$ 2200\,K,][]{Parmentier+etal+2018} that orbit very close to their host stars. The characterization of the atmospheres of UHJs has been a rapidly growing field in exoplanet science over the past five years. Theoretical studies of UHJ atmospheres suggest that the atmospheres may be dominated by atoms and ions rather than molecules due to thermal dissociation and ionization \citep{Lothringer+etal+2018,Parmentier+etal+2018,Arcangeli+etal+2018}. In addition, because their atmospheres are in an extreme state, some extreme phenomena may occur, such as atmospheric evaporation and escape \citep[e.g.,][]{YanandHenning2018,Sing+etal+2019}. Hydrodynamic atmospheric escape driven by intense stellar irradiation (extreme-ultraviolet and/or near-ultraviolet) would lead to significant mass loss, which would affect the long-term evolution of the planet \citep{Fossati+etal+2018,GarcaMuozandSchneider2019}. And UHJs are strongly irradiated and tidally locked by their host stars, which can lead to significant differences in the chemical composition and abundances of their atmospheres between the day and night sides \citep{Arcangeli+etal+2018, BellandCowan2018, Helling+etal+2019, Parmentier+etal+2018}.
	
	High-resolution transmission spectroscopy has been widely used to characterize the atmospheres of UHJs, in particular to reveal their chemical composition. For example, \citet{YanandHenning2018} found the excess H$\alpha$ absorption in KELT-9b \citep{Gaudi+etal+2017}, revealing an extended hydrogen atmosphere. \citet{Hoeijmakers+etal+2018} further reported the presence of neutral and singly ionized atomic iron (\ion{Fe}{i} and \ion{Fe}{ii}) and singly ionized atomic titanium (\ion{Ti}{ii}) in its atmosphere. \citet{Prinoth+etal+2022} presented the first unambiguous detection of TiO on an exoplanet's terminator, which plays a crucial role in the formation of thermal inversions \citep{Hubeny+etal+2003, Fortney+etal+2008}, based on transmission spectroscopy of WASP-189b. \citet{Deibert+etal+2021} and \citet{Casasayas-Barris+etal+2021} found the presence of ionized calcium (\ion{Ca}{ii}) at the terminator of WASP-76b by studying the \ion{Ca}{ii} infrared triplet ($\sim$850 nm). Neutral \ion{Fe}{i} has also been detected on this planet \citep{Ehrenreich+etal+2020}. The atmosphere of MASCARA-2b \citep[also known as KELT-20b;][]{Talens+etal+2018,Lund+etal+2017} has been studied by \citet{Casasayas-Barris+etal+2018, Casasayas-Barris+etal+2019}, and the hydrogen Balmer series (H$\alpha$, H$\beta$, and H$\gamma$) are detected along with \ion{Ca}{ii}, \ion{Fe}{ii}, and \ion{Na}{i}. Hydrogen Balmer and \ion{Ca}{ii} lines can be used as probes to study atmospheric escape and the dynamics of the upper atmosphere.
	
	WASP-33b \citep{CollierCameron+etal+2010} is a UHJ that has attracted considerable attention in recent years, orbiting a $\delta$-Scuti A-type star with a period of 1.22 days. Its host star is very bright ($V$ $\sim$ 8 mag), making it an ideal target for ground-based high-resolution observations. The dayside atmosphere of WASP-33b has been well studied. Several studies have found evidence of a temperature inversion in its dayside atmosphere \citep[e.g.,][]{Haynes+etal+2015,Nugroho+etal+2017,Cont+etal+2021}. Species such as \ion{Si}{i}, \ion{Fe}{i}, \ion{Ti}{i}, \ion{Ti}{ii}, \ion{V}{i}, OH, CO, H$_{2}$O, and TiO have been detected in the dayside atmosphere by high-resolution emission spectroscopy \citep{Nugroho+etal+2020,Nugroho+etal+2021,Serindag+etal+2021,Cont+etal+2021,Cont+etal+2022a,Cont+etal+2022b,Herman+etal+2022,vanSluijs+etal+2022,Finnerty+etal+2023}. In contrast, studies of the terminator atmosphere have lagged behind due to contamination from stellar pulsations. Its host star has a complex pulsation mode, with certain modes possibly being excited by the planet \citep{Herrero+etal+2011,Borsa+etal+2021b}. Nevertheless, the hydrogen Balmer lines and \ion{Ca}{ii} have been identified using high-resolution transmission spectroscopy \citep{Yan+etal+2021,Borsa+etal+2021b,Cauley+etal+2021,Yan+etal+2019}. In addition, \citet{vonEssen+etal+2019} claimed the detection of aluminum oxide (AlO) using a low-resolution optical transmission spectrum at the terminator of WASP-33b.
	
	To date, many ground-based high-resolution spectrographs have been used for transmission spectroscopy of the hot Jupiter and ultra-hot Jupiter atmospheres. In this paper, we present a new study of the atmosphere of WASP-33b by transmission spectroscopy using the stable high-resolution Doppler spectrograph NEID \citep{Schwab+etal+2016}, which was designed to detect exoplanets with extreme precision radial velocity and newly commissioned in late 2019. With its high spectral resolution and stability, NEID has the potential to unravel the atmospheric composition of exoplanets. In addition, we also combine previous observational data in order to obtain more robust results.
	
	This paper is organized as follows. We describe the observations and data reduction in Sect.~\ref{OBS} and~\ref{DR}. We summarize our methods, including the line-by-line analysis and the cross-correlation analysis, in Sect.~\ref{METH}. We present and discuss the resulting transmission spectra in Sect.~\ref{RandD}. The conclusions are given in Sect.~\ref{CONCL}.

\section{Observations}\label{OBS}

	We observed one transit of WASP-33b on October 16, 2021 using the NEID \citep{Schwab+etal+2016} spectrograph on the WIYN 3.5m Telescope at Kitt Peak Observatory\footnote{The data are available at \url{https://neid.ipac.caltech.edu/search.php}}. NEID covers the wavelength range from 380 to 930 nm at a resolving power of $R\sim 110,000$ or $\sim$2.7 km s$^{-1}$ \citep{Halverson+etal+2016}. The exposure time was set to 300 s. We obtained 65 spectra, but discarded four due to low signal-to-noise ratio (S/N). The average S/N of the remaining 61 spectra is around 98 at 5500 $\AA$. The raw data were reduced using the NEID Data Reduction Pipeline\footnote{\url{https://neid.ipac.caltech.edu/docs/NEID-DRP/}} (NEID-DRP). After data reduction of the raw frames, we obtained the order-by-order spectra. These spectra are at the vacuum wavelength and in the terrestrial rest frame. For the convenience of the study, we converted the wavelengths to air wavelengths. Although the NEID wavelength range covers down to 380 nm, the S/N of the bluer orders is too low to be used for transmission spectroscopy.
	
	In addition, we collected archival data for four transits of WASP-33b from the Calar Alto Archive\footnote{\url{http://caha.sdc.cab.inta-csic.es/calto/jsp/searchform.jsp}} and the Italian center for Astronomical Archive\footnote{\url{http://archives.ia2.inaf.it/tng/}}. These were used by \citet{Yan+etal+2021} to detect the hydrogen Balmer lines. Two of them were observed on January 5 and January 16, 2017, with the CARMENES \citep{Quirrenbach+etal+2018} spectrograph installed on the 3.5 m telescope at the Calar Alto Observatory. The CARMENES visual channel has a wavelength coverage of 520--960 nm with a high spectral resolution ($R\sim 94,600$). The CARMENES data were reduced using the CARACAL pipeline \citep{Caballero+etal+2016}. Similar to NEID, we also converted the wavelengths of the order-by-order one-dimensional spectra to air wavelengths. The first 19 spectra with shorter exposure times ($<$120 s) were discarded for the first night. Weather conditions were ideal on the first night and overcast on the second night.
	
	The remaining two transits were observed on October 17 and November 8, 2018, with the HARPS-North \citep[HARPS-N;][]{Cosentino+etal+2012} spectrograph on the Telescopio Nazionale Galileo (TNG). This spectrograph has a high resolution of $R\sim 115,000$ and covers the wavelength range from 383 to 690 nm. The raw data were processed using the HARPS-N pipeline (HARPS-N Data Reduction Software, version 3.7). We used the order-merged one-dimensional spectra with a wavelength step of 0.01 $\AA$ from the pipeline. Since the raw wavelength step is oversampled, we re-binned the spectrum so that the wavelength step becomes 0.03 $\AA$ \citep[as described in][]{Yan+etal+2021}. Unlike NEID and CARMENES, the HARPS-N spectra were at air wavelength and corrected for the barycentric Earth radial velocity (BERV). In terms of data quality, the second night was better than the first because the observed flux on the first night (October 17, 2018) dropped significantly when the telescope was pointed near the zenith \citep{Yan+etal+2021}.  
	
	The observation log summary, including our NEID observation and four publicly available observations, is presented in Table~\ref{ObservationInfo}. It should be noted that the CARMENES and HARPS-N observations were used only for the cross-correlation search and not for the line-by-line analysis.
	
	\begin{table*}[htb!]
	\caption{Summary of the observation logs.}             
	\label{ObservationInfo}      
	\begin{tabular}{l c c c c c c c}     
	\hline\hline 
	\noalign{\smallskip}      
	Instrument & Night & Date [UT] & Obs. Time [UT] & Exposure Time [s] & $N_\text{obs}$  & Airmass\\
	\noalign{\smallskip}
	\hline\noalign{\smallskip}                  
   		NEID & Night-1 & 2021-10-16 & 04:30$-$10:32 & 300 & 65 & 1.43-1.02-1.14\\  
	\hline\noalign{\smallskip} 
   		CARMENES & Night-1 & 2017-01-05 & 20:13$-$23:49& 120& 75 & 1.01-1.15-1.54\\
   		CARMENES & Night-2 & 2017-01-16 & 19:25$-$00:07 & 120& 66 & 1.01-1.11-2.03\\
	\hline\noalign{\smallskip} 
   		HARPS-N & Night-1 & 2018-10-17 & 21:39$-$05:46& 200& 124 & 1.64-1.01-1.54\\
   		HARPS-N & Night-2 & 2018-11-08 & 19:59$-$05:02 & 200& 141 & 1.74-1.01-1.87\\
	\noalign{\smallskip}
	\hline                  
	\end{tabular}
	\end{table*}

\section{Data reduction}\label{DR}
	
	For each NEID spectral order, we first performed a blaze correction using the blaze function provided by the NEID pipeline. We then combined all the spectral orders from each order-by-order spectrum and resampled them to a uniform grid with a wavelength step of 0.02$\AA$ to obtain an order-merged one-dimensional spectrum. To correct for outliers that are unlikely to be caused by the observed source, we flagged the pixels that were greater than ten standard deviations from the flux values of the smoothed spectrum produced by a median filter with a window size of 9 pixels. The flux values of these pixels were replaced by the median of the ten surrounding pixels \citep{Langeveld+etal+2021}. We normalized each spectrum and corrected for the continuum variation following \citet{Chen+etal+2020}. Specifically, we first averaged all the out-of-transit spectra to create a master-out spectrum as a preliminary reference spectrum. We then fitted the ratio of each individual spectrum to the reference spectrum with a fourth-order polynomial function. Finally, we divided each spectrum by the best-fit polynomial function to correct for the continuum variation. After this data reduction, we obtained the final one-dimensional normalized spectra for the following analysis.
	
	Similar to the data reduction for NEID, we obtained the order-merged normalized CARMENES spectra. We did not apply any blaze correction to these spectra as this was already done by the CARACAL pipeline. In addition, for HARPS-N, we do not need the merging operation since the HARPS-N spectrum (s1d) provided by the pipeline is one-dimensional, but the other procedures are consistent with NEID.

\section{Methods}\label{METH}

	In this section, we apply two methods commonly used in transmission spectroscopy to search for atmospheric species, the line-by-line analysis and the cross-correlation analysis. 

\subsection{Line-by-line analysis}
   
\subsubsection{Creation of transmission spectral matrix}\label{CreatTSM}
    
    We first used version 1.2.0 of \texttt{Molecfit} \citep{Smette+etal+2015,Kausch+etal+2015} to perform the telluric correction. \texttt{Molecfit} is an ESO tool for correcting telluric features in ground-based spectra. \texttt{Molecfit} produces a synthetic model of the Earth's atmospheric absorption for each spectrum based on a line-by-line radiative transfer mode (LBLRTM). Unlike CARMENES and NEID, the HARPS-N spectra are given in the Solar system barycentric rest frame. Therefore, we first shifted each HARPS-N spectrum to the terrestrial rest frame using the BERV values. The parameters used to build the telluric spectrum are similar to those of \citet{Allart+etal+2017}, but are slightly adjusted to account for instrumental differences. In addition, we carefully selected the wavelength regions containing only the separated unsaturated strong lines of telluric H$_{2}$O or O$_{2}$ to fit for the Earth's atmospheric absorption. 
	
	After the telluric correction, we shifted all the spectra to the stellar rest frame by considering the BERV values and the stellar systemic velocity ($\varv_\text{sys}$). We then computed a new master-out spectrum by combining all the continuum variation and telluric corrected out-of-transit spectra that have been corrected for continuum variation and telluric using the square of the S/N as the weight. We divided each individual spectrum by the master-out spectrum to remove the stellar lines. The spectrum in which the stellar lines are removed, called the residual spectrum, contains planetary signals and noise. We applied a Saviztky-Golay filter using \emph{savgol\_filter} from the \emph{scipy} package with a window size of 401 to remove low frequency variations on the continuum of the residual spectrum, which could be caused by stellar pulsation, blaze variation correction, or instrumental stability. Finally, we arranged all the residual spectra by phase to construct a transmission spectral matrix.
	
\subsubsection{Correction of RM and CLV effects}
	
	The residual spectra still contain the contamination from variations in the stellar line profile caused by additional systematic effects during the transit. Two important effects are the Rossiter-McLaughlin (RM) effect \citep{Rossiter1924, McLaughlin1924, Queloz+etal+2000, LoudenandWheatley2015} and the center-to-limb variation (CLV) effect \citep[e.g.,][]{Yan+etal+2015, Yan+etal+2017, Czesla+etal+2015, Cegla+etal+2016, Casasayas-Barris+etal+2018, YanandHenning2018, Chen+etal+2020}. A strong RM effect is expected in the transmission spectrum of WASP-33b because its host star is a rapidly rotating A-type star. Correction for the RM effect is therefore critical to the transmission spectroscopy study of WASP-33b. 
 
    In addition, WASP-33b's orbit is subject to nodal precession, and the orbit change rates have been measured using Doppler tomography \citep{Johnson+etal+2015, Iorio2016, Watanabe+etal+2020}. Measuring the orbital inclination ($i$) and the projected spin-orbit misalignment angle ($\lambda$) at the epoch of each observation is difficult under the effects of the stellar pulsation and is beyond the scope of this study. Therefore, based on the epoch difference between the reference epoch (2014) and each observation epoch, we adopted the orbital change rates measured by \citet{Johnson+etal+2015} to calculate the orbital inclination and spin-orbit misalignment angle at the epoch of each observation. This strategy has been used in previous studies such as \citet{Yan+etal+2019, Yan+etal+2021}. The stellar and planetary parameters of the WASP-33 system are listed in Table~\ref{WASP33INFO}.
		
	\begin{table}[htb!]
	\caption{Parameters of the WASP-33 system.}             
	\label{WASP33INFO}      
	\resizebox{\linewidth}{!}{          
	\begin{tabular}{l c c}     
	\hline\hline 
	\noalign{\smallskip}      
		Parameter & Symbol [Unit] & Value \\
	\noalign{\smallskip}
	\hline\noalign{\smallskip} 
	\multicolumn{3}{l}{\emph{Stellar parameters}} \\
	\noalign{\smallskip}                
   		Stellar mass & $M_{\star}$ $[M_{\sun}]$ & $1.561_{-0.079}^{+0.045}$ $^{(a)}$ \\  
   		\noalign{\smallskip} 
		Stellar radius & $R_{\star}$ $[R_{\sun}]$ & $1.509_{-0.027}^{+0.016}$ $^{(a)}$ \\
   		\noalign{\smallskip} 
		Effective temperature & $T_\text{eff}$ $[\text{K}]$ & $7430 \pm 100$ $^{(a)}$ \\
		\noalign{\smallskip} 
		Metallicity & $[\text{Fe/H}]$ $[\text{dex}]$ & $-0.1 \pm 0.02$ $^{(a)}$ \\
		\noalign{\smallskip} 
		Projected rotation velocity & $\varv \sin i_{\star}$ $[\text{km}$ $\text{s}^{-1}]$ & $86.63_{-0.32}^{+0.37}$ $^{(a)}$ \\
		Systemic velocity & $\varv_\text{sys}$ $[\text{km}$ $\text{s}^{-1}]$ & $-3.0 \pm 0.4$ $^{(a)}$ \\
	\noalign{\smallskip}
	\hline\noalign{\smallskip} 
	\multicolumn{3}{l}{\emph{Planetary parameters}} \\
		\noalign{\smallskip} 
		Planet mass & $M_\text{p}$ $[M_\text{J}]$ & $2.16 \pm 0.20$ $^{(a)}$ \\ 		
		\noalign{\smallskip}
		Planet radius & $R_\text{p}$ $[R_\text{J}]$ & $1.679_{-0.030}^{+0.019}$ $^{(a)}$ \\ 		
		\noalign{\smallskip}
		Planetary surface gravity & $\log g_{\text{p}}$ $[\text{cgs}]$ & $3.46_{-0.12}^{+0.08}$ $^{(b)}$ \\ 		
		\noalign{\smallskip}
		Equilibrium temperature & $T_\text{eq}$ $[\text{K}]$ & $2710 \pm 50$ $^{(a)}$ \\
		\noalign{\smallskip}
		Orbital semi-major axis & $a$ $[R_{\star}]$ & $3.69 \pm 0.05$ $^{(a)}$ \\
		\noalign{\smallskip}
		Orbital period & $P$ $[\text{d}]$ & $1.219870897$ $^{(a)}$ \\
		\noalign{\smallskip}
		Transit epoch (BJD) & $T_0$ $[\text{d}]$ & $2454163.22449$ $^{(a)}$ \\
		\noalign{\smallskip}
		Transit depth & $\delta$ $[\%]$ & $1.4$ $^{(a)}$\\
		\noalign{\smallskip}
		RV semi-amplitude & $K_\text{p}$ $[\text{km}$ $\text{s}^{-1}]$ & $231 \pm 3$ $^{(a)}$\\
		\noalign{\smallskip}
		\emph{2017 January 5}  \\
		\emph{2017 January 16} \\
		\noalign{\smallskip}
		Orbital inclination & $i$ $[\text{deg}]$ & $89.50$ $^{(c)}$ \\
		\noalign{\smallskip}
		Projected spin-orbit angle & $\lambda$ $[\text{deg}]$ & $-114.05$ $^{(c)}$ \\
		\noalign{\smallskip}
		\emph{2018 October 17} \\
		\emph{2018 November 8} \\
		\noalign{\smallskip}
		Orbital inclination & $i$ $[\text{deg}]$ & $90.14$ $^{(c)}$ \\
		\noalign{\smallskip}
		Projected spin-orbit angle & $\lambda$ $[\text{deg}]$ & $-114.93$ $^{(c)}$ \\
		\noalign{\smallskip}
		\emph{2021 October 16} \\
		\noalign{\smallskip}
		Orbital inclination & $i$ $[\text{deg}]$ & $91.19$ $^{(d)}$ \\ 
		\noalign{\smallskip}
		Projected spin-orbit angle & $\lambda$ $[\text{deg}]$ & $-116.36$ $^{(d)}$ \\
	\noalign{\smallskip}
	\hline                  
	\end{tabular}}
	\tablerefs{$^{(a)}$~\citet{Yan+etal+2019}. $^{(b)}$~\citet{Kovacs+etal+2013}. $^{(c)}$ Adopted from~\citet{Yan+etal+2019}. The changes of the orbital parameters between the two HARPS-N or two CARMENES observations are negligible, as the dates are very close. $^{(d)}$ Predicted value using parameters in~\citet{Johnson+etal+2015}.}
	\end{table}
	
	To correct for the CLV and RM effects, we followed the method described in \citet{Casasayas-Barris+etal+2019} to construct the CLV and RM model. In this method, the stellar spectra at 21 different limb-darkening angles ($\mu = \cos \theta$) are computed with the \texttt{Spectroscopy Made Easy} (SME) tool \citep{ValentiandPiskunov1996, PiskunovandValenti2017}, using the VALD3 line lists \citep{Ryabchikova+etal+2015} and the Kurucz ATLAS9 models \citep{CastelliandKurucz2003}. The stellar spectrum calculation assumes local thermodynamic equilibrium (LTE) and solar abundance. However, we found that the synthetic stellar spectrum is slightly different from the observational data. In this case, we fitted the abundances of some crucial elements (e.g., H for the analysis of H$\alpha$, H$\beta$, and H$\gamma$ lines) using the SME tool instead of directly assuming the solar abundance. We roughly assumed that the size of the planetary disk was equal to the apparent radius of the planet, and did not account for the increase in radius due to the planetary atmosphere. Using this method, we obtained a combined CLV and RM model and normalized it by creating a master-out CLV+RM model, similar to the creation of the master-out spectrum in Sect.~\ref{CreatTSM}.
	
\subsection{Cross-correlation analysis}

\subsubsection{Spectral models}\label{SM}

	We created a model spectrum for each species to be analyzed using the petitRADTRANS code \citep{Molliere+etal+2019}. This code is widely used to calculate synthetic transmission spectra of planetary atmospheres of ultra-hot Jupiters \citep[e.g.,][]{Hoeijmakers+etal+2019, Stangret+etal+2020, Borsa+etal+2021a, Sedaghati+etal+2021, Stangret+etal+2021, Kesseli+etal+2022, Casasayas-Barris+etal+2022}.
	
	For the case of WASP-33b, we assumed a solar abundance as input for each species and an isothermal atmosphere at 2700 K, which is close to the equilibrium temperature of WASP-33b. The planetary radius and surface gravity we used are listed in Table~\ref{WASP33INFO}. The line lists of the species used to compute the spectral models, except for TiO, are provided by petitRADTRANS, including AlO \citep{Patrascu+etal+2015}, FeH \citep{Wende+etal+2010}, VO (B. Plez) and other atoms and ions available in the Kurucz \citep{Kurucz+etal+2017} database, which have been used in previous studies \citep[e.g.,][]{Casasayas-Barris+etal+2022}. For TiO, we chose to use the TOTO line list, calculated by \citet{McKemmish+etal+2019} and stored in the ExoMol database, because it is relatively accurate \citep{Serindag+etal+2021}. According to \citet{Hoeijmakers+etal+2019}, the atmosphere becomes opaque between 1 and 10~mbar due to H$^{-}$ absorption in ultra-hot Jupiters. Therefore, according to the nature of WASP-33b, we set the continuum level to 10~mbar.
		 	
	Before the spectral models were used in the cross-correlation analysis, they were convolved with a Gaussian function to match the instrument resolution. The spectral models were normalized with a Gaussian high-pass filter of 12 points to remove the continuum level. With these procedures, the spectral models have values around one and are ready for use in the cross-correlation analysis.
	
\subsubsection{Cross-correlation}\label{DR_CC}

	For efficiency and simplicity, we used the SYSREM algorithm \citep{Tamuz+etal+2005}, which is widely used in cross-correlation analysis, to reduce the telluric and stellar contamination and obtain the observed transmission spectrum. The cross-correlation analysis is different from the line-by-line analysis, but the two methods produce consistent results. 
	
	We first constructed the spectral matrix by collecting the one-dimensional normalized spectra of each observation after data reduction (see Sect.~\ref{DR}), and then ran SYSREM on the spectral matrix. In one iteration, SYSREM captures the systematics by iteratively fitting each wavelength bin in the spectral matrix, resulting from variations in telluric absorption, stellar lines, and instrumental effects. SYSREM then removes these systematics from the spectral matrix. Through multiple iterations, SYSREM can effectively remove the telluric and stellar lines and obtain the residual transmission spectral matrix containing the planetary features and noise. We implemented SYSREM following \citet{Gibson+etal+2020} by dividing the spectral matrix by the systematics rather than subtracting the systematics from the spectral matrix. We ran the algorithm through ten iterations and selected the residual spectral matrix of the iteration with the highest S/N as the algorithm output. This matrix was then used in the cross-correlation calculations.
	
	We implemented the cross-correlation analysis \citep{Snellen+etal+2010} for each species independently. The spectral models were shifted from $-$250 to $+$250 km~s$^{-1}$ with a step of 1 km~s$^{-1}$. At each shift, the weighted residual spectrum was multiplied by the shifted spectral model to obtain a weighted cross-correlation function (CCF). The CCF is given by
   \begin{equation}
      \mathrm{CCF}=\sum \frac{r_{i} m_{i}(\varv)}{\sigma_{i}^{2}},
   \end{equation}
   where $r_{i}$ is the residual spectrum, $m_{i}$ is the spectral model shifted by velocity $\varv$, and $\sigma_{i}$ is the error at wavelength point $i$. We calculated the variance of each wavelength bin (i.e., along the column) and the variance of each frame (i.e., along the row) in the residual spectral matrix, and then took the square root of their sum as the error. For each residual spectrum, we obtained a CCF. We then stacked all the CCFs to generate the CCF map for each observation night and spectral model.

	To detect planetary signals, we used a grid of assumed planetary orbital velocity semi-amplitudes ($K_\text{p}$) to convert the CCF map from the terrestrial rest frame to the planetary rest frame. Assuming the planet has a circular orbit, the radial velocity (RV) of the planet $\varv_\text{p}$ is defined as
   \begin{equation}
      \varv_{\mathrm{p}}=\varv_{\text {sys }}+\varv_{\text {bary }}+K_{\mathrm{p}} \sin (2 \pi \phi)+\Delta \varv,
   \end{equation}
   where $\varv_\text{sys}$ is the systemic velocity, $\varv_\text{bary}$ is the barycentric Earth radial velocity (i.e., BERV), $\Delta \varv$ is the radial velocity deviation from zero value, and $\phi$ is the orbital phase. In general, $\Delta \varv$ can be closely related to the dynamical properties of the planetary atmosphere (e.g., wind). For each value of the assumed $K_\text{p}$, we generated a one-dimensional CCF by averaging the CCF map of the in-transit portion, which had been shifted to the planetary rest frame by that value. We then created a $K_\text{p}$ map by stacking the one-dimensional CCFs of the different $K_\text{p}$ values ranging from 0 to 300 km~s$^{-1}$ with a step of 1 km~s$^{-1}$. The $K_\text{p}$ map was expressed in the form of S/N, with the CCF values being normalized by the standard deviation calculated from $+$250 to $+$300 km~s$^{-1}$ in $K_\text{p}$ and $+$200 to $+$250 km~s$^{-1}$ in $\Delta \varv$.
   
\section{Results and discussion}\label{RandD}

\subsection{Species detection using the line-by-line technique}
	We present here the results obtained by analyzing individual lines in different wavelength regions. Using the NEID data, we confirm previous detections of the Balmer series of H (H$\alpha$, H$\beta$, and H$\gamma$) and \ion{Ca}{ii}. We also try to analyze the spectral lines of several atoms and ions expected to be present in WASP-33b's atmosphere. However, since almost all individual lines of atoms and ions are affected by stellar pulsations, we cannot distinguish between planetary atmospheric absorption signals and stellar pulsation signals. Therefore, we could not report any additional new significant detections of atoms and ions in WASP-33b using line-by-line analysis. While in principle a large number of additional transit observations or pulsation modeling with simultaneous (spectro-)photometric monitoring can help to eliminate the effects of pulsations, we decided to treat stellar pulsations as correlated noise in our analysis, as will be detailed in this section. 

\subsubsection{H$\alpha$ transmission spectrum}\label{SectHalpha}

   \begin{figure}
   \centering
   \includegraphics[width=\hsize]{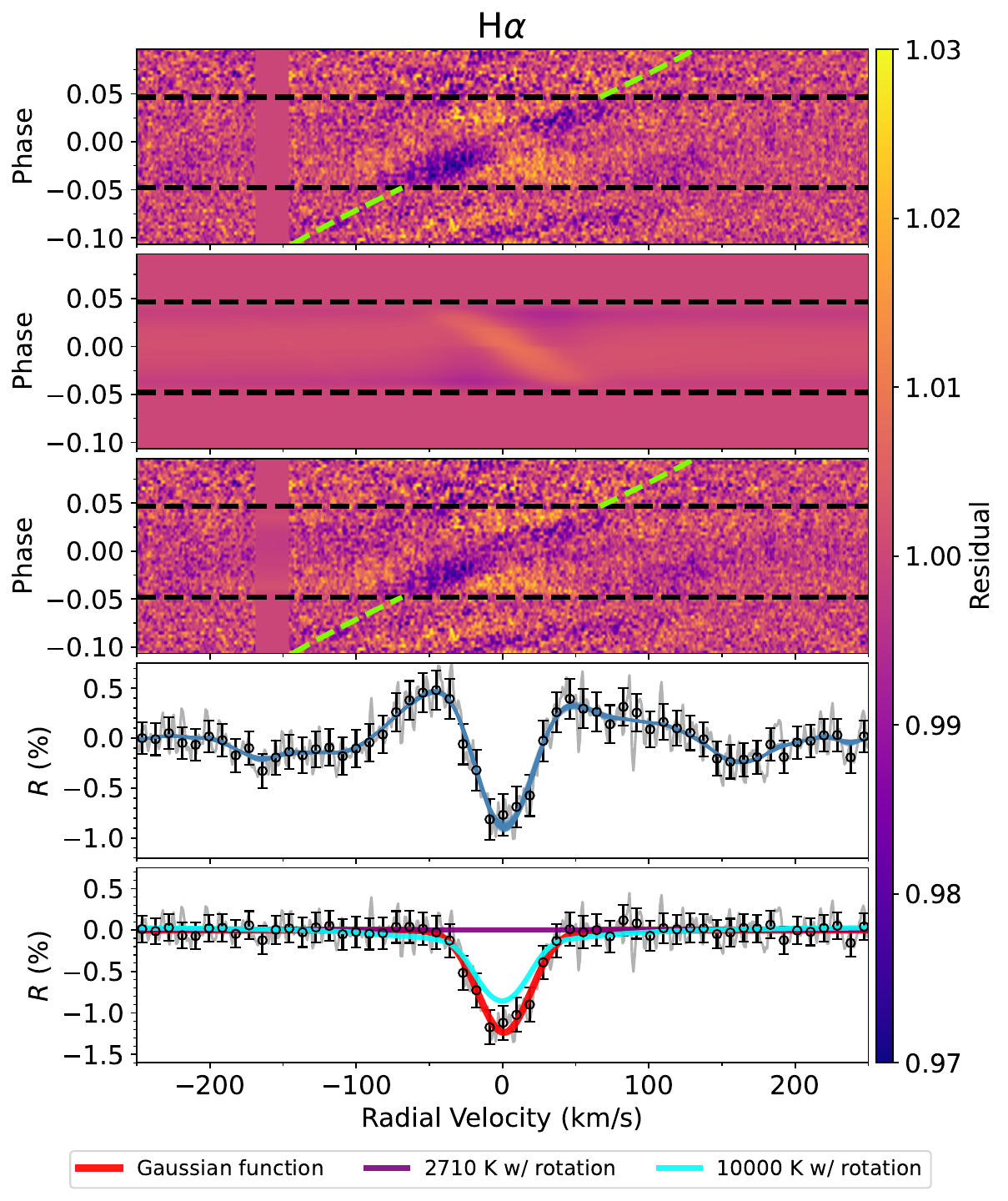}
   \caption{Transmission spectral matrix, CLV+RM model, and one-dimensional transmission spectrum for the H$\alpha$ line observed with NEID. \emph{First row}: the transmission spectral matrix before correction for the CLV and RM effects in the stellar rest frame. The horizontal black dashed lines indicate the first and fourth contacts of the transit (i.e., the beginning and end of the transit). The slanted green dashed lines mark the radial velocity shift due to the orbital motion of the planet. The radial velocity regions around $-158$ km~s$^{-1}$ (i.e. $\sim$ 6559.35 \AA) are masked due to invalid values (NaN). \emph{Second row}: the model for the CLV and RM effects. \emph{Third row}: the observed transmission spectral matrix after correction for the CLV and RM effects. The color bar is indexed to the residual flux. \emph{Fourth row}: phase-folded one-dimensional transmission spectrum. The CLV and RM effects have been corrected. The $y$-axis shows the relative residual flux minus one in $\%$. The gray line shows the unbinned observed transmission spectrum. The black circles with error bars show the spectrum binned every ten points ($\sim$0.2 $\AA$). The blue line shows the best-fit GP model. \emph{Fifth row}: transmission spectrum after removal of systematic noise using the GP method. The red line shows the best-fit Gaussian line profile. The purple and cyan lines are hydrostatic models of H$\alpha$ with temperatures of 2,710 K and 10,000 K, respectively, including planetary rotation.}
   \label{NEID2DHalpha}%
   \end{figure}

	The first to third rows of Figure~\ref{NEID2DHalpha} shows the transmission spectral matrix and the stellar CLV and RM effects around the H$\alpha$ line. The planetary H$\alpha$ absorption is clearly visible in the phase-resolved transmission spectra observed with NEID after correction for the CLV and RM effects. The velocity shifts of the absorption are fully consistent with the orbital motion of the planet during the transit. We removed the CLV and RM effects from the transmission spectral matrix, and shifted the residual spectra to the planetary rest frame using the expected $K_{\text{p}}$ = 231 km~s$^{-1}$ calculated with the planetary orbital parameters. We then averaged all the in-transit spectra with the square of the S/N as weight to derive the final one-dimensional transmission spectrum. The phase-folded H$\alpha$ transmission spectrum is shown in the fourth row of Figure~\ref{NEID2DHalpha}. We fit a Gaussian profile to the one-dimensional H$\alpha$ transmission spectrum, assuming that the planetary absorption signal has a Gaussian profile controlled by the line contrast ($h$), the velocity offset of the center of the observed line ($V_\text{center}$), and the full width at half maximum (FWHM). The best-fit parameters derived by this classical white-noise based method are presented in Table~\ref{NEID1DFIT}.
	
	\begin{table*}[htb!]
	\caption{The best-fit parameters from the classical and GP methods fit to the one-dimensional transmission spectra.}             
	\label{NEID1DFIT}      
	\centering          
	\begin{tabular}{c l c c c c c c}     
	\hline\hline 
	\noalign{\smallskip}      
	 & & \multicolumn{3}{c}{Classical method} & \multicolumn{3}{c}{GP method}\\
	\cmidrule(r){3-5} \cmidrule(r){6-8}
	Line & Instrument & $V_\text{center}$ (km~s$^{-1}$) & $h$ (\%) & \emph{FWHM} (km~s$^{-1}$) & $V_\text{center}$ (km~s$^{-1}$) & $h$ (\%) & \emph{FWHM} (km~s$^{-1}$) \\
	\noalign{\smallskip}
	\hline\noalign{\smallskip}                  
   		H$\alpha$ & NEID & 1.9 $\pm$ 0.7 & 0.91 $\pm$ 0.04  & 30.5 $\pm$ 1.7 & 1.1 $^{+1.6}_{-1.6}$ & 1.24 $^{+0.20}_{-0.19}$ & 42.8 $^{+3.1}_{-3.2}$\\
	\noalign{\smallskip}
   		 & CARMENES \tablenotemark{\footnotesize *} &0.8$\pm$1.1 & 1.11 $\pm$ 0.07  & 35.6 $^{+2.2}_{-2.0}$ & 0.4 $^{+1.2}_{-1.2}$ & 1.11 $^{+0.07}_{-0.07}$ & 36.1 $^{+2.4}_{-2.3}$\\
	\noalign{\smallskip}
   		 & HARPS-N \tablenotemark{\footnotesize *} &2.0 $\pm$ 1.9 & 0.81 $\pm$ 0.09  & 31.6 $^{+4.1}_{-3.6}$ & 1.8 $^{+2.0}_{-2.1}$ & 0.79 $^{+0.10}_{-0.10}$ & 31.0 $^{+4.4}_{-3.9}$\\ 
	\noalign{\smallskip}
	\hline\noalign{\smallskip} 
		H$\beta$ & NEID & 0.9 $\pm$ 1.1 & 0.63 $\pm$ 0.04   & 32.3 $\pm$ 2.6 & 0.2 $^{+3.2}_{-3.1}$ & 0.62 $^{+0.11}_{-0.12}$ & 36.5 $^{+5.7}_{-5.2}$\\
	\noalign{\smallskip}
		 & HARPS-N \tablenotemark{\footnotesize *} &2.2 $\pm$ 1.7 & 0.54 $\pm$ 0.07   & 30.6 $^{+4.9}_{-4.3}$ & 1.4 $^{+2.8}_{-3.0}$ & 0.52 $^{+0.10}_{-0.11}$ & 31.3 $^{+8.1}_{-6.9}$\\
	\noalign{\smallskip}
	\hline\noalign{\smallskip} 
   		H$\gamma$ & NEID & 10.5 $\pm$ 2.5 & 0.48 $\pm$ 0.07 & 33.4 $\pm$ 5.8 & 11.6 $^{+4.9}_{-5.9}$ & 0.46 $^{+0.15}_{-0.16}$ & 33.0 $^{+11.7}_{-11.7}$\\
	\noalign{\smallskip}
		 & HARPS-N \tablenotemark{\footnotesize *} &7 $\pm$ 15 & 0.28 $^{+0.09}_{-0.15}$ & 49 $^{+25}_{-26}$ & $-$2.6 $^{+10.8}_{-7.2}$ & 0.29 $^{+0.17}_{-0.19}$ & 36.2 $^{+32.7}_{-18.1}$\\ 
	\noalign{\smallskip} 
	\hline\noalign{\smallskip} 
   		\ion{Ca}{ii} IRT $(\boldsymbol{\lambda}8498)$ & NEID & 3.8 $\pm$ 1.2 & 0.38 $\pm$ 0.03 & 28.2 $\pm$ 2.9 & 1.9 $^{+3.0}_{-3.5}$ & 0.40 $^{+0.10}_{-0.10}$ & 35.6 $^{+8.7}_{-8.0}$\\
	\noalign{\smallskip} 
   		\ion{Ca}{ii} IRT $(\boldsymbol{\lambda}8542)$ & NEID & 4.3 $\pm$ 0.7 & 0.71 $\pm$ 0.05    & 21.9 $\pm$ 1.7 & 2.8 $^{+1.9}_{-2.1}$ & 0.78 $^{+0.15}_{-0.15}$ & 30.7 $^{+5.9}_{-5.4}$\\
	\noalign{\smallskip} 
		\ion{Ca}{ii}~IRT $(\boldsymbol{\lambda}8662)$ & NEID & $-$1.6 $\pm$ 0.6 & 0.78 $\pm$ 0.05    & 21.0 $\pm$ 1.5 & $-1.8$ $^{+1.5}_{-1.5}$ & 0.81 $^{+0.12}_{-0.12}$ & 25.3 $^{+4.9}_{-4.0}$\\
	\noalign{\smallskip} 
	\hline\noalign{\smallskip} 
   		\ion{Ca}{ii} IRT & CARMENES \tablenotemark{\footnotesize *} &2.0 $\pm$ 0.7 & 0.67 $\pm$ 0.04 & 26.0 $\pm$ 1.7 & 1.9 $^{+1.4}_{-1.4}$ & 0.67 $^{+0.07}_{-0.08}$ & 26.2 $^{+3.8}_{-3.4}$\\
	\noalign{\smallskip}
	\hline                  
	\end{tabular}
	\tablecomments{\footnotesize * The results of the classical method of CARMENES and HARPS-N are obtained from \citet{Yan+etal+2019,Yan+etal+2021}.}
	\end{table*} 
	
	The center of the H$\alpha$ line measured with NEID has a velocity offset of $1.9 \pm 0.7 $ km~s$^{-1}$. Velocity offsets inferred by transmission spectroscopy could be due to planetary winds \citep{Wyttenbach+etal+2015, LoudenandWheatley2015, Brogi+etal+2016}. Based on HARPS-N measurements, \citet{Borsa+etal+2021b} reported a significant blueshift of $-$8.2 $\pm$ 1.4 km~s$^{-1}$ for H$\alpha$ and attributed it to the presence of winds in the atmosphere of WASP-33b. However, \citet{Yan+etal+2021} found no evidence for winds (velocity offset of $1.2 \pm 0.9 $ km~s$^{-1}$) by combining HARPS-N and CARMENES observations. Our results are more consistent with the latter. In the case of WASP-33b, these contradictory results are not surprising because of the correlated noise introduced by stellar pulsations. The pulsations would temporarily alter the stellar line profile and make it difficult to remove the stellar lines using the master-out spectrum, leaving strong residuals on top of the planetary absorption. Such correlated noise could introduce spurious velocity offsets when fitting a Gaussian function to the expected planetary absorption line. 
	
    Meanwhile, because its host star is a rapidly rotating A-type star, it is extremely difficult to accurately measure the stellar systemic RV. Without a precise RV value, we are cautious about interpreting the measured velocity offset as the winds in the planetary atmosphere. We conclude that our NEID data do not show a significant velocity shift, but we cannot exclude the possibility that winds exist at the terminator of WASP-33b.
	
	The FWHM of the H$\alpha$ line is measured to be $30.5 \pm 1.7$ km~s$^{-1}$, which is slightly lower than the results of \citet{Borsa+etal+2021b} and \citet{Yan+etal+2021}. The difference could be due to instrumental effects, different data reduction procedures (e.g., removal of telluric and stellar lines, correction of CLV and RM effects, and normalization), and stellar pulsations. Given the visible variations of stellar pulsations on different nights, stellar pulsations could be the more likely origin. Such a large FWHM is the result of a combination of mechanisms, including rotational broadening, thermal broadening, and dynamical processes, suggesting that excited hydrogen atoms may be present in the upper atmosphere and likely undergo hydrodynamic escape.

	The measured line contrast of the H$\alpha$ line is $0.91 \pm 0.04$ \%. Following \citet{Chen+etal+2020}, we derived the effective planetary radius ($R_\text{eff}$) defined by: $R_\text{eff}^2/R_\text{p}^2 = (\delta+h)/\delta$, where $\delta = (R_\text{p}/R_\text{$\star$})^2$ is obtained from Table~\ref{WASP33INFO} and $h$ is the measured line contrast. For the H$\alpha$ line, the effective planetary radius is $1.302_{-0.012}^{+0.012}$ $R_\text{p}$, which is below the effective Roche radius ($1.71_{-0.07}^{+0.08} R_\text{p}$ for WASP-33b). Therefore, the observed H$\alpha$ absorption signal originates from the gravitationally bound atmospheric envelope within the Roche lobe of the planet. Consequently, we will not be able to observe out-of-transit absorption like \citet{Nortmann+etal+2018} and \citet{Ehrenreich+etal+2015} when planetary atmospheric materials do not escape beyond the Roche lobe. We note that our measured line contrast is consistent with \citet{Yan+etal+2021} ($0.99 \pm 0.05$ \%), but is almost twice the value obtained by \citet{Borsa+etal+2021b} ($0.54 \pm 0.04$ \%) and significantly lower than a third measurement ($1.68 \pm 0.02$ \%) reported by \citet{Cauley+etal+2021} using the PEPSI spectrograph mounted on the Large Binocular Telescope. As with the FWHM, we speculate that the discrepancies in the measured line contrast are caused by different levels of stellar pulsations. 

	In order to obtain proper parameter estimates for the line profile, we decided to adopt Gaussian process (GP) regression in the line profile modeling. GP can capture the correlation between data points based on a covariance matrix, where data points that are close to each other in input space are highly correlated. Since its first implementation in low-resolution transmission spectroscopy by \citet{Gibson+etal+2012}, GP has been widely used in time-series exoplanet data to account for correlated noise, and has recently been proposed to treat telluric and planetary signals for high-resolution transmission spectroscopy \citep{Meech+etal+2022}. Here we used GP to account for the correlated systematics caused by stellar pulsations.

	We assumed that the one-dimensional transmission spectrum ($R$) follows a multivariate Gaussian distribution:
   \begin{equation}
      R \sim \mathcal{GP}\left(M(\lambda ; \theta), C(\lambda ; \varphi)\right),
   \end{equation}
   where $M$ is the mean function with the parameter vector $\theta$, $C$ is the covariance matrix with the parameter vector $\varphi$, and $\lambda$ is the wavelengths. The elements in $C$ are determined by a kernel function $k\left(\tau_{i j} ; \varphi\right) \text {, where } \tau_{i j} \equiv \left|\lambda_{i}-\lambda_{j}\right|$ is the wavelength interval between two data points. Here we assumed that the mean function $M$ is the Gaussian function to model a pure planetary absorption profile. We implemented GP with the python code \emph{celerite} \citep{Foreman-Mackey+etal+2017}. The kernel function consists of a 3/2-order Mat\'ern kernel and a diagonal jitter ("white noise") term. Finally, we used the Markov chain Monte Carlo (MCMC) sampling algorithm implemented in \emph{emcee} \citep{Foreman-Mackey+etal+2013} to find the best-fit parameters and their uncertainties. The best-fit parameters and adopted priors are presented in Table~\ref{NEID1DFIT} and Table~\ref{NEID1DFITGPPARAM}, respectively.  
	
	The blue line in the fourth row of Figure~\ref{NEID2DHalpha} shows the best-fit model derived from the GP method, where the GP does a good job of accounting for the variation caused by stellar pulsations. The GP method derived an FWHM of $42.8_{-3.2}^{+3.1}$ km~s$^{-1}$, a line contrast $h$ of $1.24_{-0.19}^{+0.20}$ \% ($R_\text{eff}$ = 1.396$^{+0.054}_{-0.053}$ $R_\text{p}$), and a velocity offset of $1.1_{-1.6}^{+1.6}$ km~s$^{-1}$. Compared to those derived from the classical white-noise based method, the GP-derived parameters are more robust with larger error bars after accounting for the correlated noise. 
	
	The purple and cyan lines in the fifth row of Figure~\ref{NEID2DHalpha} show the hydrostatic models of H$\alpha$ with 2,710 K (i.e., the equilibrium temperature of WASP-33b) and 10,000 K, respectively. The planetary rotation broadening is considered assuming tidal locking using the method of \citet{Boucher+etal+2023}. Each hydrostatic model is calculated by petitRADTRANS assuming an isothermal temperature structure, chemical equilibrium, and other settings consistent with Sect.~\ref{SM}. The observed H$\alpha$ excess absorption is significantly stronger than the equilibrium-temperature model prediction and slightly stronger than the 10,000~K model prediction. We note that the maximum temperature allowed by our adopted H line opacity is 6,000~K, which might lead to the underestimation. On the other hand, the discrepancy could be due to the lack of consideration of non-local thermodynamic equilibrium (NLTE) effects and/or hydrodynamic escape \citep[e.g.,][]{GarcaMuozandSchneider2019,Fossati+etal+2021,Fossati+etal+2023,Huang+etal+2023}. 
 	
	We also applied the GP method to the H$\alpha$ transmission spectra derived in \citet{Yan+etal+2021}. As shown in Table~\ref{NEID1DFIT}, the results obtained by the classical method and the GP method are very consistent for the H$\alpha$ transmission spectra of \citet{Yan+etal+2021}. This is probably because the line core of the planetary absorption profile is less affected by stellar pulsations in the corresponding observations with different phasing. 
	
   \begin{figure}
   \centering
   \includegraphics[width=\hsize]{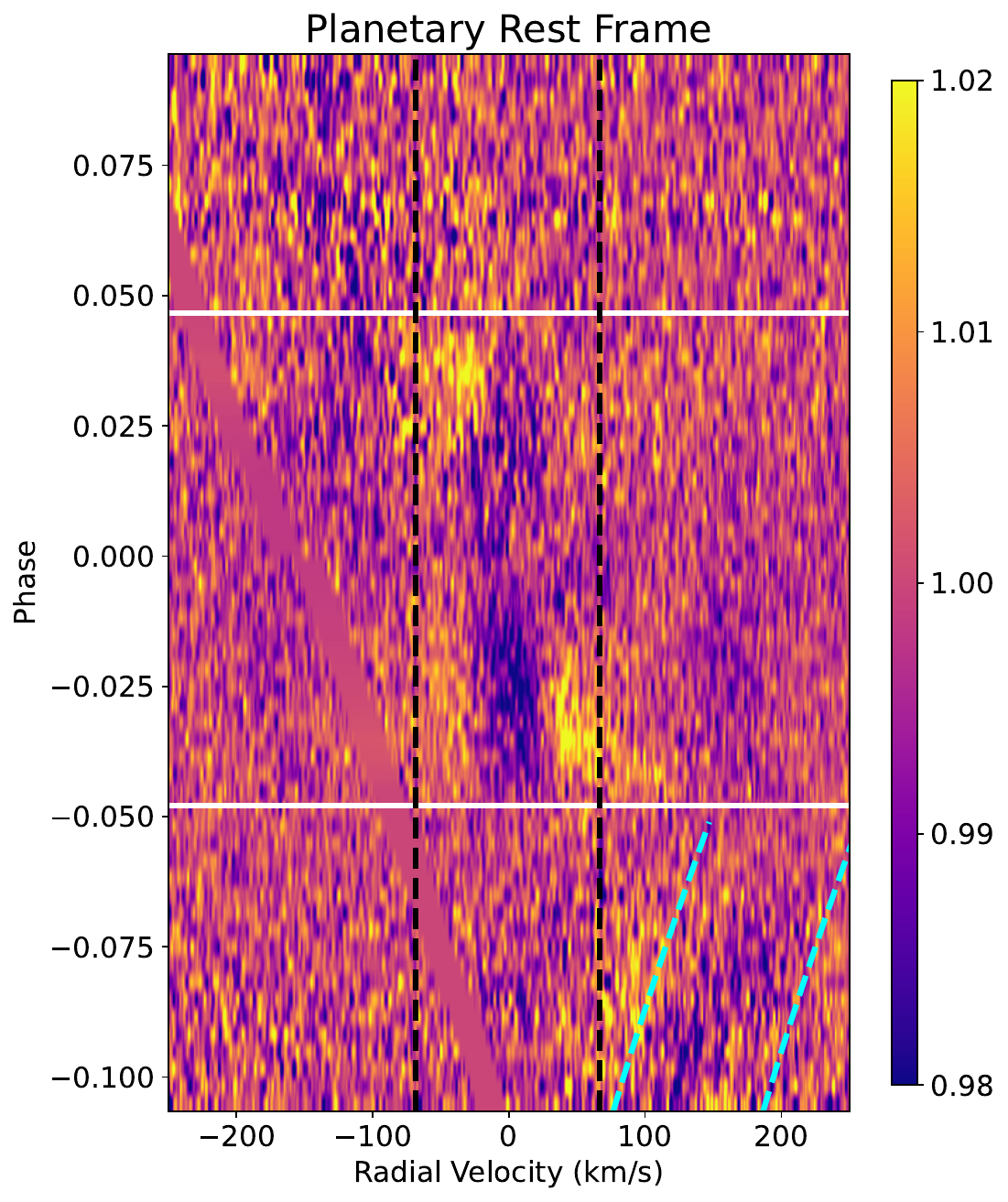}
   \caption{Transmission spectral matrix of the H$\alpha$ absorption in the planetary rest frame. The horizontal white lines indicate the first and fourth contacts of the transit. The vertical black dashed lines represent the values of the planetary radial velocity at the first and fourth contacts of the transit. The slanted cyan dashed lines show the radial velocity trace of the planet moved to bracket the pre-transit absorption feature. The color bar indicates the value of residual flux.}
   \label{NEID1DHalpha_PRF}
   \end{figure}
	
	We note that there is an absorption feature moving at a velocity semi-amplitude of $\sim$460~km~s$^{-1}$ (approaching twice the expected $K_{\text{p}}$ of the planet) in the pre-transit phase, as shown at the third panel of Figure~\ref{NEID2DHalpha}. When shifted to the planetary rest frame, as shown in Figure~\ref{NEID1DHalpha_PRF}, the pre-transit absorption feature is located at red-shifted velocities of more than 100~km~s$^{-1}$. Similar pre-transit absorption features have been reported by \citet{Borsa+etal+2021b} and \citet{Cauley+etal+2021}. \citet{Borsa+etal+2021b} found that the pre-transit absorption features did not cancel out like other pulsations when averaging different transits, and suggested that they could be stellar pulsation mode excited by the planet. The evidence of tidally perturbed stellar pulsation modes has also been found in the light curve of WASP-33 observed by the Transiting Exoplanet Survey Satellite, which were attributed to the misalignment of the stellar rotational axis and the planetary orbit \citep{Kalman+etal+2022}.

\subsubsection{H$\beta$ and H$\gamma$ transmission spectra}\label{SectHbetaHgamma}
	
\begin{figure*}[htb!]
   \centering
   \includegraphics[width=\hsize]{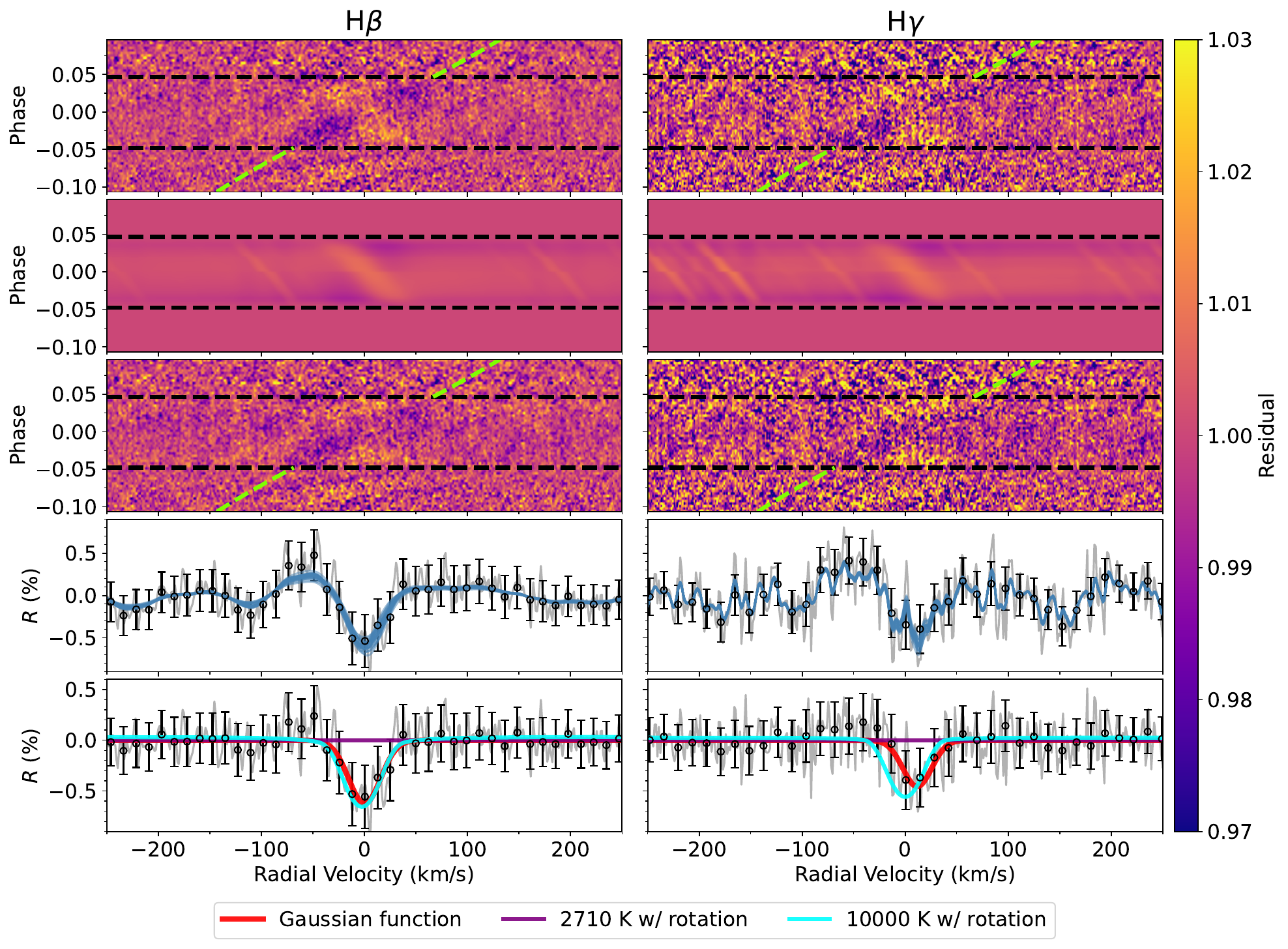}
   \caption{The transmission spectral matrices and one-dimensional transmission spectra for the H$\beta$ and H$\gamma$ lines observed by NEID. Each column corresponds to an individual line. The \emph{first row} to the \emph{fifth row} are similar to the forms of Figure~\ref{NEID2DHalpha}.}
   \label{NEIDHbetagamma}
   \end{figure*}

	The NEID spectrograph also covers the H$\beta$ and H$\gamma$ lines. The best-fit parameters of the Gaussian function using the classical white-noise based method and the GP method are shown in Table~\ref{NEID1DFIT}. The H$\beta$ and H$\gamma$ transmission spectral matrices and one-dimensional transmission spectra are shown in Figure~\ref{NEIDHbetagamma}. For the H$\beta$ and H$\gamma$ lines, the NEID data results are in agreement with those of \citet{Yan+etal+2021} measured on the HARPS-N data. The effective radius of the H$\beta$ and H$\gamma$ lines are 1.214$^{+0.034}_{-0.038}$ $R_\text{p}$, and 1.163$^{+0.048}_{-0.054}$ $R_\text{p}$, respectively. The velocity shifts of the H$\beta$ and H$\gamma$ lines are consistent with planetary orbital motion, but their absorption features are weaker than the H$\alpha$ line. Moreover, their absorption depths are significantly stronger than hydrostatic model predictions at the equilibrium temperature. The case of the hydrogen Balmer series has also been detected in KELT-9b \citep{Cauley+etal+2019, Wyttenbach+etal+2020} and KELT-20b \citep{Casasayas-Barris+etal+2019}.
	
\subsubsection{\ion{Ca}{ii} transmission spectra}\label{CaIILBL}
	Although the \ion{Ca}{ii} H\&K lines (K line 3933.661 $\AA$, and H line 3968.467 $\AA$ at air wavelength) induced by resonant transitions from the \ion{Ca}{ii} ground state are covered by the wavelength region of the NEID spectrograph, the S/N of these lines is too low to extract the absorption signals from the planetary atmospheric layer. Therefore, we only analyze the three near-infrared triplet (IRT) lines with the NEID data in this section.
	
\begin{figure*}[htb!]
   \centering
   \includegraphics[width=\hsize]{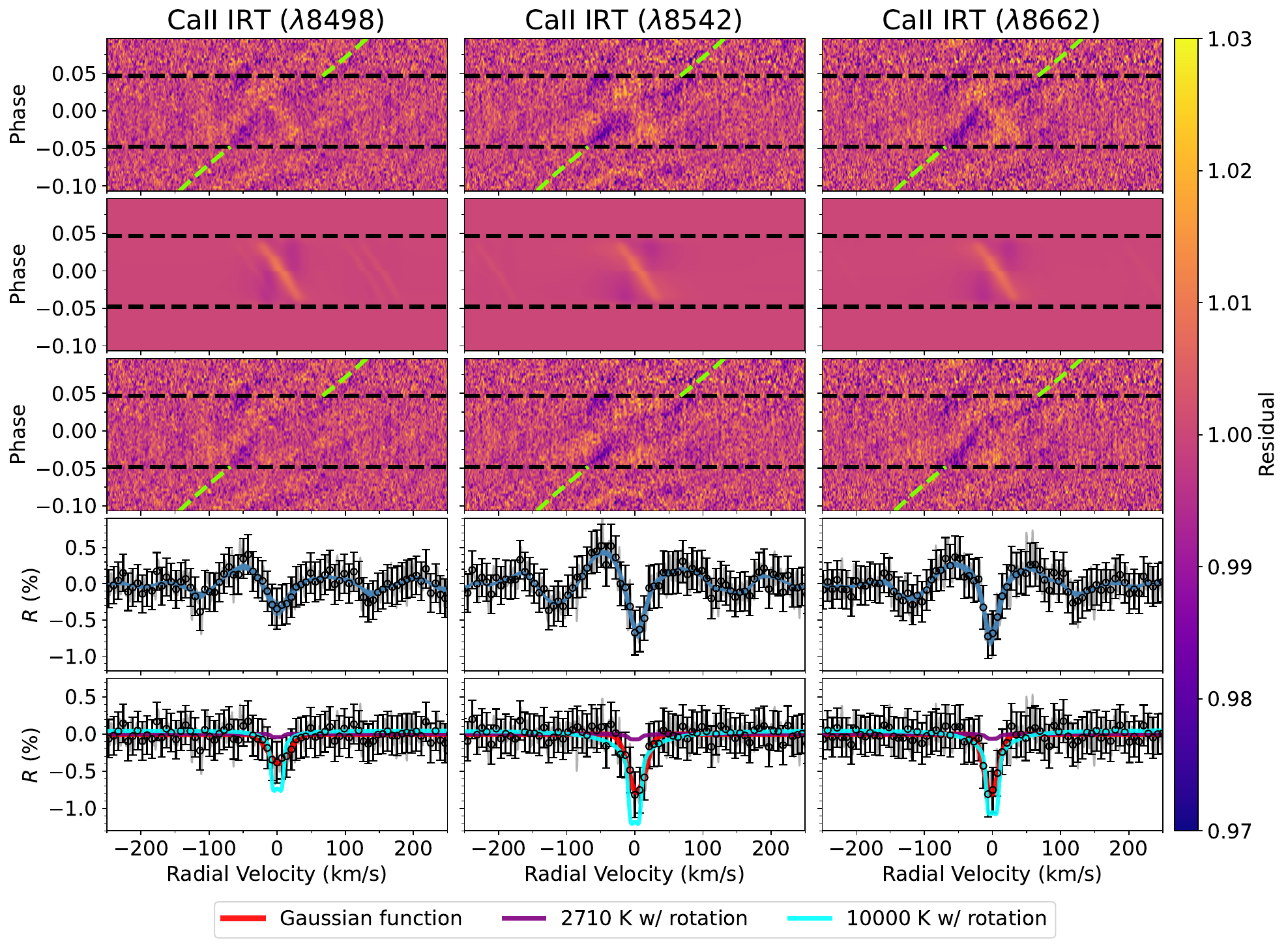}
   \caption{As in Figure~\ref{NEIDHbetagamma}, but for the \ion{Ca}{ii} IRT lines, including \ion{Ca}{ii} IRT $(\boldsymbol{\lambda}8498)$, \ion{Ca}{ii} IRT $(\boldsymbol{\lambda}8542)$, and \ion{Ca}{ii}~IRT $(\boldsymbol{\lambda}8662)$. The name of each line is given in the title of each top panel.}
   \label{NEIDCaIRT}
   \end{figure*}

	Similar to Sect.~\ref{SectHbetaHgamma}, the transmission spectral matrix and one-dimensional transmission spectrum for each line of the \ion{Ca}{ii} IRT (8498.018 $\AA$, 8542.089 $\AA$, and 8662.140 $\AA$ at air wavelength, labeled as \ion{Ca}{ii} IRT $\boldsymbol{\lambda}8498$, \ion{Ca}{ii} IRT $\boldsymbol{\lambda}8542$, and \ion{Ca}{ii}~IRT $\boldsymbol{\lambda}8662$, respectively) are shown in Figure~\ref{NEIDCaIRT} from left to right. Table~\ref{NEID1DFIT} shows the best-fit parameters of the Gaussian function based on the classical method and the GP method, respectively. The mean velocity offset of the \ion{Ca}{ii} lines is $0.51_{-1.32}^{+1.36}$ km~s$^{-1}$ for the GP method. This result is consistent with the Balmer lines when stellar pulsations are treated as correlated noise by the GP, indicating that no strong day- to night-side winds are observed. The mean FWHM of the \ion{Ca}{ii} IRT lines ($29.7_{-3.3}^{+3.6}$ km~s$^{-1}$) is consistent with previous results. We translate the line contrast of \ion{Ca}{ii} into an effective planetary radius of $R_\text{eff} = 1.143_{-0.034}^{+0.033}$ $R_\text{p}$ for \ion{Ca}{ii} IRT $(\boldsymbol{\lambda}8498)$, $R_\text{eff} = 1.264_{-0.046}^{+0.045}$ $R_\text{p}$ for \ion{Ca}{ii} IRT $(\boldsymbol{\lambda}8542)$, and $R_\text{eff} = 1.273_{-0.037}^{+0.036}$ $R_\text{p}$ for \ion{Ca}{ii}~IRT $(\boldsymbol{\lambda}8662)$, respectively. In addition, similar to the Balmer lines, the \ion{Ca}{ii} absorption is also significantly stronger than predicted by the equilibrium temperature hydrostatic model. This is consistent with \citet{Yan+etal+2019}.

\subsection{Species detection using the cross-correlation technique}

	We used the cross-correlation technique to search for some species that are expected to be present at the terminator of WASP-33b, in particular TiO, \ion{Ti}{i}, \ion{V}{i}. These have been detected in the dayside atmosphere of WASP-33b. In general,  materials from the planet's dayside atmosphere can be transported through the terminator to the nightside by certain dynamical processes. Thus, if species are detected in the dayside atmosphere, we are possibly to find them in the terminator and nightside atmosphere. In addition, we looked for some other atoms and ions that are expected to be present in ultra-hot Jupiter. Unfortunately, these atoms and ions are affected by the pulsations of WASP-33, and we have difficulty distinguishing between pulsation signals and planetary atmospheric absorption signals.
	
   \begin{figure}[htb!]
   \centering
   \includegraphics[width=\hsize]{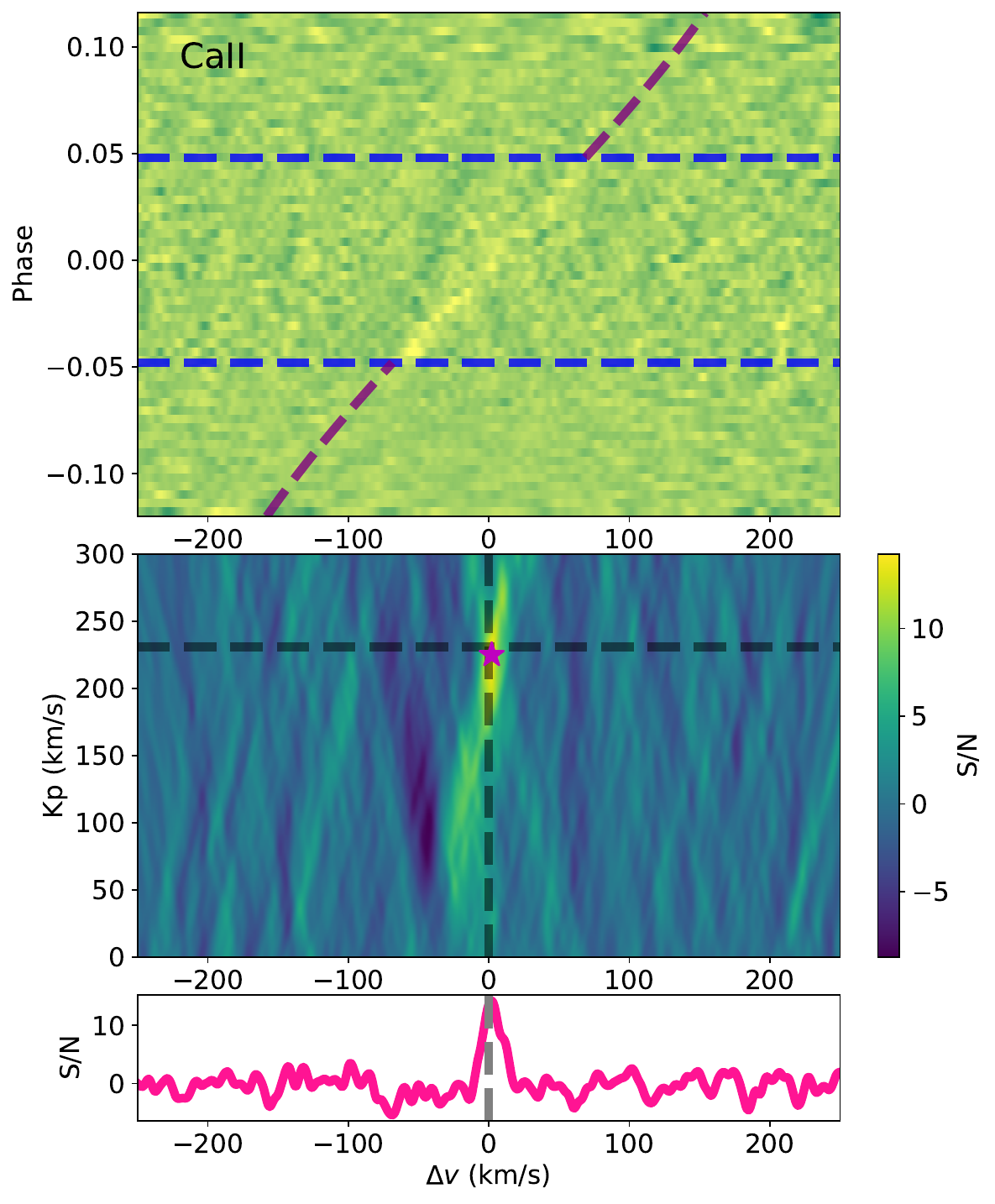}
   \caption{Cross-correlation functions of \ion{Ca}{ii} averaged over the five nights of observation. \emph{First row}: the CCF map of \ion{Ca}{ii} after correction for CLV and RM effects. The horizontal blue dashed lines represent the first and fourth contacts of the transit. The purple dashed line shows the orbital motion of the planet. \emph{Second row}: $K_{\text{p}}$ map. The magenta pentagram marks the S/N peak. The expected values are indicated by the dashed lines. The color bar represents the value of the S/N. \emph{Third row}: the one-dimensional cross-correlation function (the pink dashed line) corresponding to the $K_{\text{p}}$ value at S/N peak. The vertical gray dashed line indicates $\Delta \varv$ = 0.}
   \label{CCFKPCaII}
   \end{figure}

\subsubsection{Presence of \ion{Ca}{ii} confirmed in CCF}	

	In Sect.~\ref{CaIILBL}, we confirmed the presence of \ion{Ca}{ii} in the atmosphere of WASP-33b using the line-by-line technique with the NEID data. Here we combined the five observations (see Table~\ref{ObservationInfo}) to obtain the CCF and $K_{\text{p}}$-maps for \ion{Ca}{ii}. To reduce the CLV+RM effects, we applied the CLV+RM model to correct each residual spectrum before cross-correlation. The corrected maps are shown in Figure~\ref{CCFKPCaII}.
	
   \begin{figure*}[htb!]
   \centering
   \includegraphics[width=\hsize]{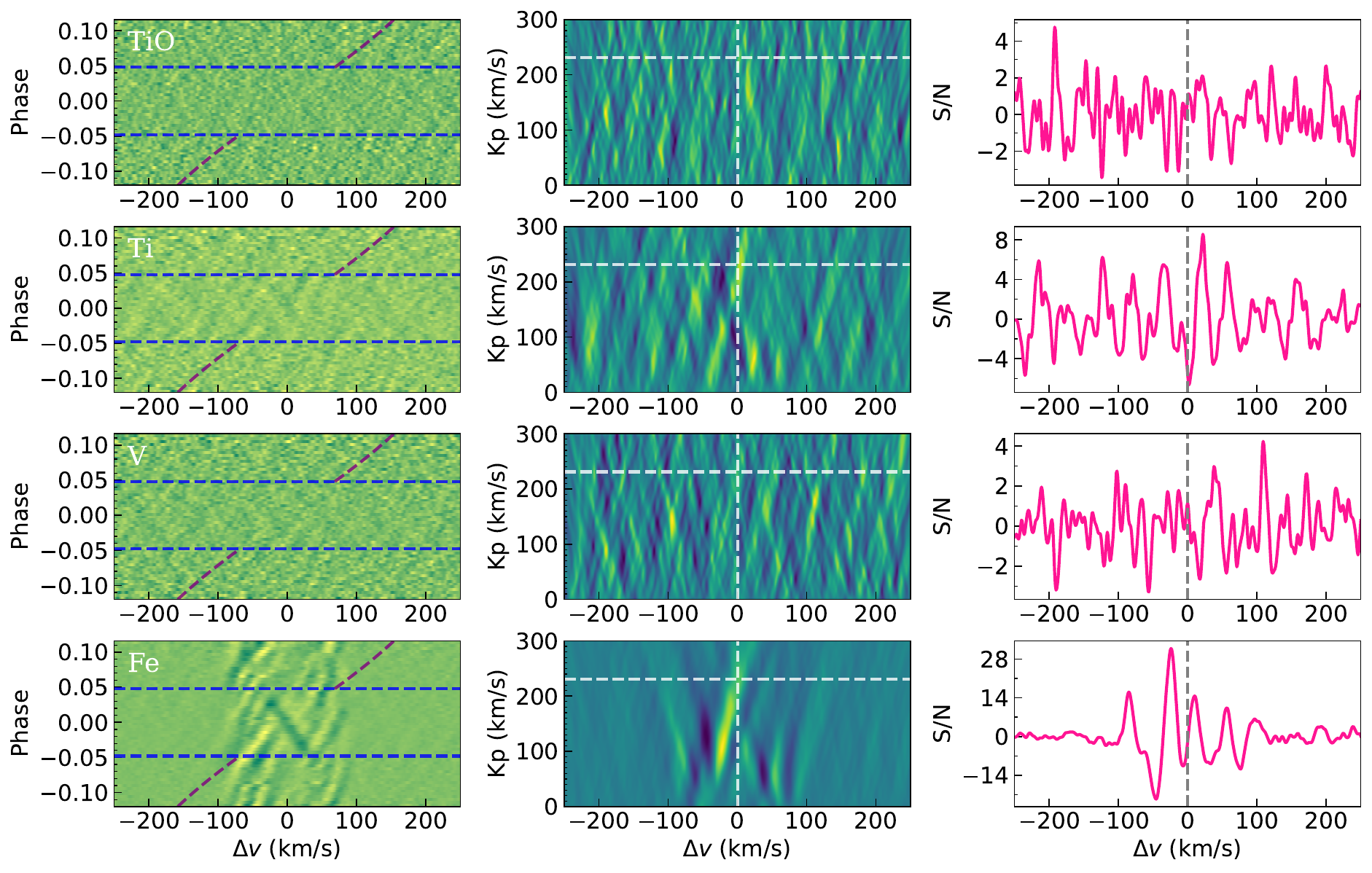}
   \caption{Cross-correlation functions of TiO, \ion{Ti}{i}, \ion{V}{i}, and \ion{Fe}{i} without correction for CLV and RM effects. Each row of panels represents one species. The format of each panel from \emph{left} to \emph{right} corresponds to each row from top to bottom of Figure~\ref{CCFKPCaII}, respectively. The name of each species is given in each \emph{top-left panel}.}
   \label{CCFKP}
   \end{figure*}
    
	For WASP-33b, the \ion{Ca}{ii} signal is very strong, and we can observe it directly from the CCF map (see the top panel of Figure~\ref{CCFKPCaII}). The $K_{\text{p}}$ map is shown in the middle panel of Figure~\ref{CCFKPCaII}. This map shows a strong cross-correlation signal at the expected values with an S/N of 14.2. The bottom panel of Figure~\ref{CCFKPCaII} shows the CCFs at the $K_{\text{p}} = 225$ km~s$^{-1}$, corresponding to the S/N peak. The peak of the CCFs located at $\Delta \varv \sim 0$ km~s$^{-1}$ indicates the absence of wind in the atmosphere of WASP-33b.
	
	We derived a $K_{\text{p}}$ value of $225_{-21}^{+11}$ km~s$^{-1}$ using the \ion{Ca}{ii} absorption lines, including the H\&K lines and the IRT lines. The derived $K_{\text{p}}$ value using planetary \ion{Ca}{ii} absorption is in agreement with the expected $K_{\text{p}}$ value \citep[$231_{-3}^{+3}$ km~s$^{-1}$, taken from][]{Yan+etal+2019} calculated using Kepler's third law. \citet{Yan+etal+2019} reported a similar $K_{\text{p}}$ ($= 224$ km~s$^{-1}$) value using the \ion{Ca}{ii} lines at the terminator of WASP-33b. \citet{Cont+etal+2022b} obtained a $K_{\text{p}}$ of $225_{-2.5}^{+3.0}$ km~s$^{-1}$ using the planetary emission lines (including the lines of \ion{Ti}{i}, \ion{V}{i}, OH, \ion{Fe}{i}, \ion{Si}{i}, and \ion{Ti}{ii}) in WASP-33b and speculated that the discrepancy between the derived value and the expected value may be due to the super-rotation of WASP-33b. 
	
	In addition, the pulsations of WASP-33b's host star also affect the constraint on the $K_\text{p}$ value. This is because pulsations could change the stellar line profile \citep{CollierCameron+etal+2010} and induce spurious signals similar to planetary absorption in the CCF map. These spurious signals could overlap with the signals of planetary origin. When the pulsation component dominates, the weak planetary signal is overwhelmed by noise, making it difficult to detect species in the planetary atmosphere. For \ion{Ca}{ii}, the strong cross-correlation signal is located at the planetary velocity and appears only during transit, it is unlikely to be induced by pulsation. However, some species with weak features in WASP-33b's atmosphere or with high abundances in WASP-33b's host star are difficult to detect at the terminator using transmission spectroscopy, especially for metal atoms and ions, as will be discussed in the following text.

\subsubsection{Non-detection of TiO}

   \begin{figure}
   \centering
   \includegraphics[width=\hsize]{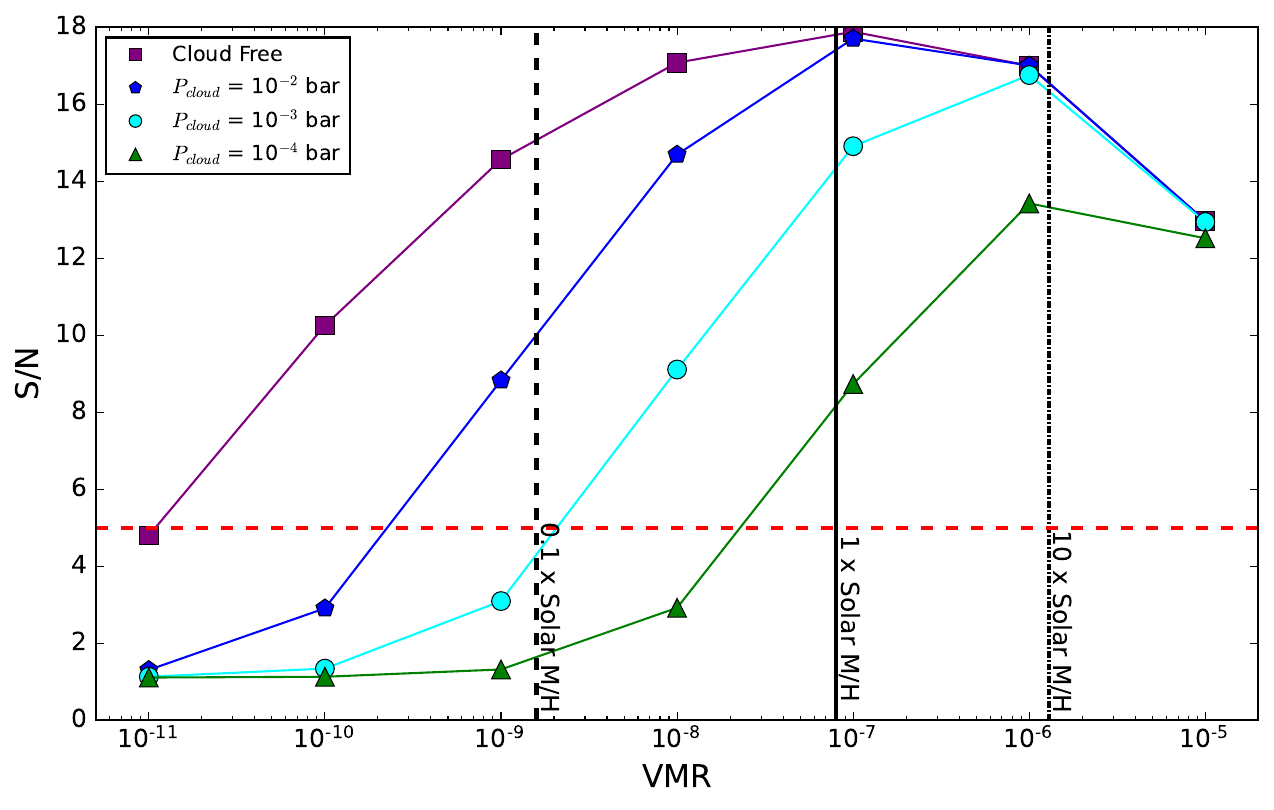}
   \caption{Summary of injection-recovery tests for TiO. The horizontal red dashed line indicates S/N = 5. The three vertical lines represent the VMR of TiO at 1 bar from the equilibrium chemical calculations for different metallicities. We assume that 1 bar is the quenching point, and the the VMR of TiO is constant above this point. The y-axis indicates the S/N at the expected values in the $K_\text{p}$ map.}
   \label{TiO_injection}
   \end{figure}
   
	The detection of TiO is not expected to be confounded by stellar pulsations. Due to the extremely high temperature of  the host star's atmosphere, molecules such as TiO could not be present. Therefore, we can search for TiO by applying the cross-correlation technique in transmission spectroscopy. 
	
	We combined the five observations to search for the presence of TiO at the terminator of WASP-33b. The cross-correlation results of TiO are shown in the first row of Figure~\ref{CCFKP}. There are no planetary features as obvious as \ion{Ca}{ii} in the CCF map of TiO, and an S/N peak is not located at the expected value ($K_{\text{p}} = \text{231 km~s}^{-1}$, $\Delta \varv = \text{0 km~s}^{-1}$). We conclude that the peak is likely due to noise and that there is no evidence for the presence of TiO at the terminator of WASP-33b in the data we used. 
	
	We performed injection-recovery tests for TiO. We injected models with different volume mixing ratios (VMRs) of TiO (see Sect.~\ref{SM}) into the data, and reduced the data after injection and performed the cross-correlation analyses according to Sect.~\ref{DR_CC}. We injected models with the VMR of TiO from 10$^{-11}$ to 10$^{-5}$ spaced by 1 dex. We also considered some cloudy scenarios (i.e., $P_\text{cloud}$ = $10^{-2}$, $10^{-3}$, and $10^{-4}$ bar). In general, the S/N $\textgreater$ 5 at the expected values in the $K_\text{p}$ map is used as a threshold for robust detection. We used FastChem \citep{Stock+etal+2018} to calculate the VMRs of TiO from the equilibrium chemical models with the planet's equilibrium temperature and different metallicities. Figure~\ref{TiO_injection} shows the results of the injection-recovery tests for TiO. We recovered the injection signal at the minimum VMR of TiO of 10$^{-10}$ assuming cloud free. However, we did not find TiO in the real data, which means that the true value of the VMR of TiO at the terminator of WASP-33b is even lower and is much lower than expected from the chemical equilibrium calculations. Even in the presence of high-altitude clouds, the upper limit derived from the injection-recovery tests is lower than the VMR of TiO expected for WASP-33b with its high metallicity \citep[2$-$15$\times$ solar,][]{Finnerty+etal+2023}. Overall, the possibility of TiO not being detected at the terminator owing to the presence of high-altitude clouds and/or the influence of H$^-$ (corresponding to $P_\text{cloud}$ = 10$^{-2}$ bar) can be roughly ruled out. However, TiO has been detected in the thermal emission spectrum of WASP-33b \citep{Nugroho+etal+2017, Cont+etal+2021}, there should be certain mechanisms to allow its presence on the dayside but absence at the terminator.
	
	The lack or low abundance of TiO at the terminator is still an unsolved puzzle. Several mechanisms have been proposed to explain the absence of TiO at the terminator. The colder temperature on the nightside could cause condensation or settling of TiO in the deeper atmosphere, while the strong day-night winds on ultra-hot Jupiters may prevent efficient transport of TiO to the dayside \citep{Hubeny+etal+2003,Spiegel+etal+2009,Parmentier+etal+2013,Parmentier+etal+2016}. TiO could be strongly photo-dissociated due to the high UV flux of the host star, especially during the active cycle of the star \citep{Knutson+etal+2010}. TiO could also be strongly thermally dissociated on the dayside for ultra-hot Jupiters, and the detection of TiO at the terminator would depend on its recombination timescale relative to the atmospheric circulation timescale \citep{Lothringer+etal+2018,Parmentier+etal+2018}. The detection of TiO on the dayside of WASP-33b and the absence of TiO at the terminator in our results seems to indicate that TiO is condensed at the terminator (and the nightside), but is transported to the dayside by the atmospheric circulation and converted to the gas phase for detection. Further theoretical modelling and observations are needed to confirm this.
	
\subsubsection{Results of \ion{Ti}{i}, \ion{V}{i}, and other atomic and molecular species}
	
	\citet{Cont+etal+2022b} reported the first detection of the emission signature of \ion{Ti}{i} and \ion{V}{i} in a dayside of an exoplanet atmosphere and found no residual stellar pulsation features present in the CCF maps for \ion{Ti}{i} and \ion{V}{i} in WASP-33b. The absence of stellar pulsation is conducive to the detection of \ion{Ti}{i} and \ion{V}{i}. In our case, we tried to combine five observations to detect \ion{Ti}{i} and \ion{V}{i} in the transmission spectrum. The results are shown in the second and third rows of Figure~\ref{CCFKP}.
	
	Although the $K_\text{p}$ map of \ion{Ti}{i} shows a suspicious signal ($K_{\text{p}} \text{ = } 247_{-34}^{+7} \text{km~s}^{-1}$, $\Delta \varv \text{ = } 3_{-2}^{+2} \text{km~s}^{-1}$) at an S/N of 6.9, which is close to the expected value, there is no significant cross-correlation signal originating from the planetary atmosphere during the transit in the CCF map of \ion{Ti}{i}. The suspicious signal could also be due to a weak stellar pulsation pattern in the CCF map. Therefore, we are cautious in concluding that \ion{Ti}{i} is present at the terminator of WASP-33b. For \ion{V}{i},there is no evidence for its presence at the terminator of WASP-33b from the data we used, in agreement with the results of \citet{Borsa+etal+2021b}, who found no evidence for \ion{V}{i} using other HARPS-N observations. 
	
	A large number of \ion{Ti}{i} and \ion{V}{i} lines are covered in the wavelength range by the data we used. Thus, the non-detection of \ion{Ti}{i} and \ion{V}{i} is unlikely to be caused by a low number of spectral lines. The significant detection of \ion{Ti}{i} and \ion{V}{i} on the dayside, but not on the terminator, could be caused by the difference in the atmospheric temperature between the dayside and the terminator. Moreover, some observations suggest that temperature is not the only factor, as other physical parameters also influence the presence of refractory species such as Ti and V \citep{Cont+etal+2022b}.

	For completeness, we also searched for other atomic and molecular species using the cross-correlation technique, including \ion{Fe}{i}, \ion{Fe}{ii}, \ion{Ca}{i}, \ion{Mg}{i}, \ion{Cr}{i}, \ion{Y}{i}, VO, AlO, and FeH. The fourth row of Figure~\ref{CCFKP} illustrates the CCF results of \ion{Fe}{i} as a typical example of stellar pulsation. The CCF results for the other atomic and molecular species are shown in Figure~\ref{appendix2}. Almost all CCF maps of the atomic species are dominated by the stellar pulsation patterns and the RM effect. In the stellar rest frame, the affected velocity range is determined by the rotational velocity of the host star, about $\pm$87 km~s$^{-1}$ for WASP-33. This range just overlaps with the radial velocity of the planet during transit. The CCF maps show that although the pulsation is different from the orbital velocity semi-amplitude of the planetary motion, the strong cross-correlation signal of the pulsation can completely obscure the planetary signal. As a result, the $K_\text{p}$ maps are strongly dominated by stellar pulsations.

	In contrast, the CCF maps of the molecular species are free of the stellar pulsation patterns and the RM effect because they are not present in the high-temperature atmosphere of the host star. As for TiO, we find no evidence for VO, FeH, and AlO. However, a spurious signal ($K_{\text{p}} = 208_{-15}^{+14} \text{km~s}^{-1}$ , $\Delta \varv = -4_{-2}^{+3} \text{km~s}^{-1}$) at an S/N of 2.89 in the $K_\text{p}$ map of AlO cannot be ruled out. Further high-resolution observations are required to confirm the inference of AlO based on the low-resolution transmission spectrum \citep{vonEssen+etal+2019}. 

\section{Conclusions}\label{CONCL}
	
	We have observed a single transit of the ultra-hot Jupiter WASP-33b with the NEID spectrograph. With the newly acquired NEID data, we confirmed the previous detections of the Balmer series of hydrogen (H$\alpha$, H$\beta$, and H$\gamma$) and the \ion{Ca}{ii} infrared triplet using the line-by-line technique. We introduced the Gaussian process method to account for the correlated systematics caused by stellar pulsations when fitting the line profiles of the Balmer lines and the \ion{Ca}{ii} triplet. As a result, we robustly derived a line contrast of $1.24_{-0.19}^{+0.20}$ \% for H$\alpha$, $0.62_{-0.12}^{+0.11}$ \% for H$\beta$, $0.46_{-0.16}^{+0.15}$ \% for H$\gamma$, $0.40_{-0.10}^{+0.10}$ \% for \ion{Ca}{ii} $\lambda 8498$, $0.78_{-0.15}^{+0.15}$ \% for \ion{Ca}{ii} $\lambda 8542$, $0.81_{-0.12}^{+0.12}$ \% for \ion{Ca}{ii} $\lambda 8662$ from the NEID data. We did not find a significant net Doppler shift traced by the Balmer lines or the \ion{Ca}{ii} infrared triplet, indicating no evidence for a day- to night-side wind in WASP-33b.
    
    \citet{YanandHenning2018} detected 1.15$\pm$0.05~\% of the H$\alpha$ absorption in the atmosphere of KELT-9b, which orbits an A0-type star. The absorption corresponds to a hydrogen atmosphere about 1.64 times the radius of the planet, close to the Roche lobe (1.91$_{-0.26}^{+0.22} R_\text{p}$). \citet{Casasayas-Barris+etal+2018,Casasayas-Barris+etal+2019} claimed that the detection of Balmer series of H and \ion{Ca}{ii} in the ultra-hot Jupiter KELT-20b, orbiting an A2-type star, implies an extended atmosphere produced by irradiation far from the Roche lobe of this planet. \citet{Fossati+etal+2023} argued that KELT-9b and KELT-20b host hydrostatic atmospheres, and the Balmer series of H and \ion{Ca}{ii} lines do not probe the exosphere. This is because their host stars are early A-type and do not have strong X-ray and EUV emission to turn a hydrostatic atmosphere into hydrodynamics. Given an A5-type for WASP-33, the upper atmosphere of WASP-33b could be strongly heated to the hydrodynamic regime \citep{Fossati+etal+2018}. In addition, \citet{Fossati+etal+2023} shows that it is necessary to consider NLTE effects in the atmospheric modelling of the Balmer series of H and \ion{Ca}{ii}. Overall, the absorption of \ion{Ca}{ii} and/or Balmer series of H can be linked to the state of the planet's upper atmosphere, but is highly model dependent and requires consideration of hydrodynamic and/or NLTE effects. More detailed atmospheric modelling and comparative studies between planets with different spectral types of host stars are needed.
	
	Using the cross-correlation technique and combining the NEID data with the archival data (HARPS-N and CARMENES), we further confirmed that the \ion{Ca}{ii} absorption is from the planet, since it is consistent with the orbital motion of the planet. However, we found no evidence for the species expected to be present at the terminator, such as TiO, \ion{Ti}{i}, and \ion{V}{i}. These have already been detected in the dayside atmosphere and are not affected by stellar pulsation. The discrepancy between the species inferred from the dayside and terminator atmospheres can be attributed to factors such as differences in abundance, temperature, data reduction procedure, and systematics, which require further high-quality observations to reconcile. In addition, we looked for other atomic species and found them difficult to detect in the transmission spectrum because they are strongly affected by stellar pulsation. 
	
	So far, a number of ultra-hot Jupiters have been found. Due to the extreme temperature difference between the dayside and nightside atmospheres, the chemical composition and physical properties of day and night are expected to be significantly different \citep{Parmentier+etal+2018}, requiring more phase-resolved, high-precision observations. For WASP-33b, its dayside has been well studied, but its terminator has been poorly studied due to the impact of stellar pulsation. A detailed modeling of the stellar pulsation may be necessary for future transmission spectroscopy of WASP-33b to robustly recover the signals of planetary origin.

\begin{acknowledgments}

    The authors thank the anonymous reviewer for the constructive comments and suggestions on the manuscript. 
    G.C. acknowledges the support by the National Natural Science Foundation of China (grant Nos. 12122308, 42075122), the B-type Strategic Priority Program of the Chinese Academy of Sciences (grant No. XDB41000000), Youth Innovation Promotion Association CAS (2021315), and the Minor Planet Foundation of the Purple Mountain Observatory. 
    This paper contains data taken with the NEID instrument, which was funded by the NASA-NSF Exoplanet Observational Research (NN-EXPLORE) partnership and built by Pennsylvania State University. NEID is installed on the WIYN telescope, which is operated by the National Optical Astronomy Observatory, and the NEID archive is operated by the NASA Exoplanet Science Institute at the California Institute of Technology. NN-EXPLORE is managed by the Jet Propulsion Laboratory, California Institute of Technology under contract with the National Aeronautics and Space Administration.
    This research used the facilities of the Italian Center for Astronomical Archive (IA2) operated by INAF at the Astronomical Observatory of Trieste. 
    Based on data from the CAHA Archive at CAB (INTA-CSIC). The CAHA Archive is part of the Spanish Virtual Observatory project funded by MCIN/AEI/10.13039/501100011033 through grant PID2020-112949GB-I00". 
\end{acknowledgments}

\bibliography{manuscript.bib}{}
\bibliographystyle{aasjournal}

\appendix

\section{Additional tables and figures}

\begin{figure*}[htb!]
\centering
\includegraphics[width=\hsize]{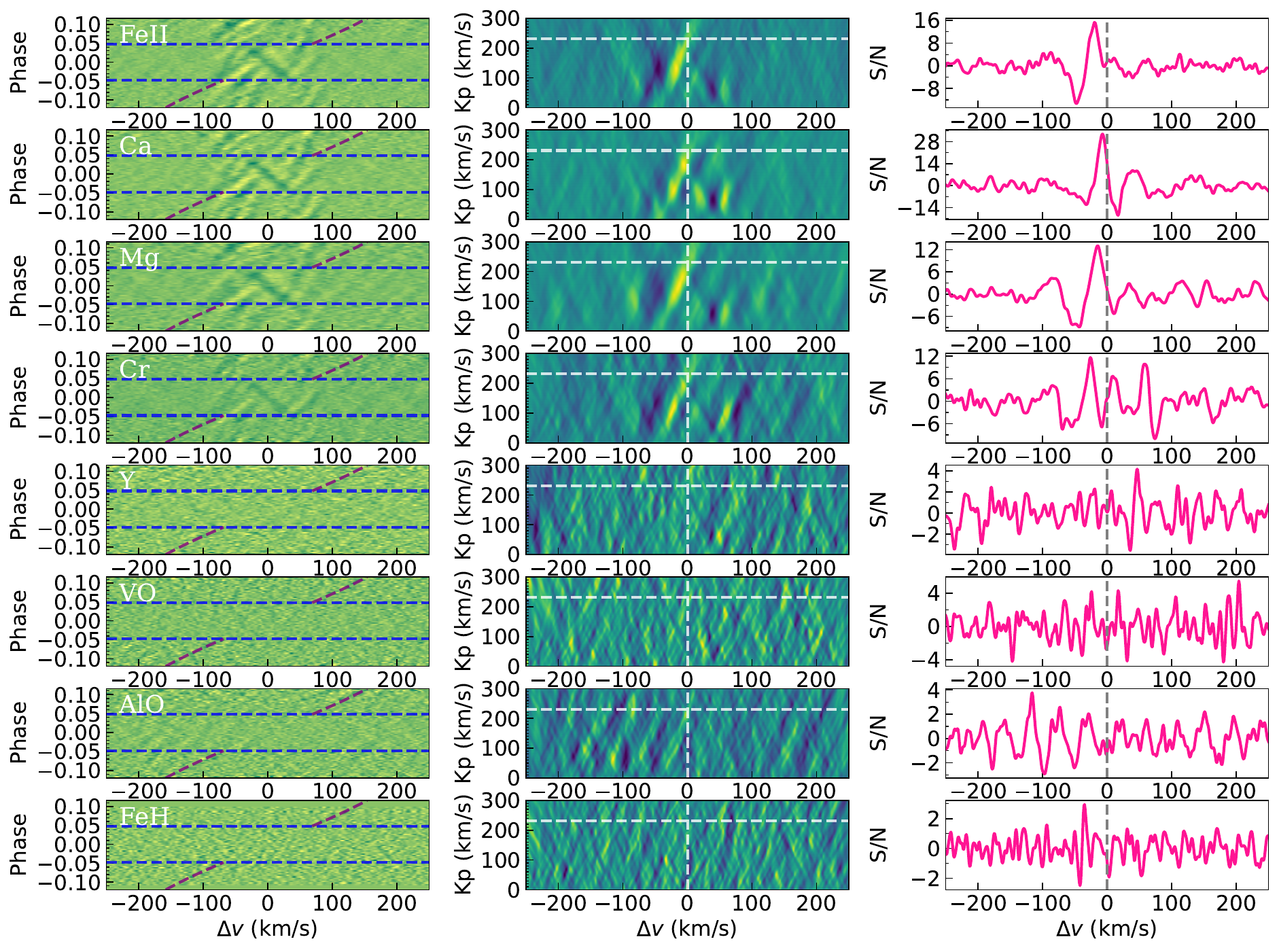}
\caption{Cross-correlation functions of the remaining searched species (\ion{Fe}{ii}, \ion{Ca}{i}, \ion{Mg}{i}, \ion{Cr}{i}, \ion{Y}{i}, VO, AlO, and FeH).}
\label{appendix2}
\end{figure*}

	\begin{table*}[htb!]
	\caption{Adopted priors for the free parameters in the GP method.}             
	\label{NEID1DFITGPPARAM}      
	\centering          
	\begin{tabular}{l c}     
	\hline\hline     
	Parameter & Prior\\
	\hline\noalign{\smallskip} 
		$h$ [\%] & $\mathcal{U}(-2,2)$ \\ 
	\noalign{\smallskip}                 
   		$\lambda_\text{center}$ [\AA] & $\mathcal{U}(\lambda_0-0.2,\lambda_0+0.2)\tablenotemark{\footnotesize a}$ \\  
	\noalign{\smallskip} 
   		$\sigma$ [\AA] & $\mathcal{U}(0.0,1.0)$ \\ 
	\noalign{\smallskip} 
   		$\ln A$ [\%] & $\mathcal{U}(-10,-1)$ \\
	\noalign{\smallskip} 
   		$\ln\tau$ [\AA] & $\mathcal{U}(-8,8)$ \\
	\noalign{\smallskip} 
		$\ln\sigma_w$ [\AA] & $\mathcal{U}(-10,0)$ \\
	\noalign{\smallskip} 
	\hline                  
	\end{tabular}
	\tablenotetext{\footnotesize a}{$\lambda_0$ represents the theoretical wavelength of each individual line.}
	\end{table*}

\end{document}